\documentclass[journal=jacsat,manuscript=article]{achemso}
\usepackage[version=3]{mhchem} 
\usepackage{siunitx}
\usepackage[dvipsnames]{xcolor}
\usepackage{amssymb}
\usepackage{pifont}
\usepackage{wrapfig}
\usepackage{hyperref}

\usepackage{amsmath} 

\usepackage{tabularx,colortbl}
\usepackage{capt-of}
\usepackage{adjustbox}

\author{Seongwon Kim}
\affiliation
{Department of Chemical Engineering, Carnegie Mellon University, 15213, USA}

\author{Parisa Mollaei}
\affiliation[meche]
{Department of Mechanical Engineering, Carnegie Mellon University, 15213, USA}

\author{Amir Barati Farimani}
\affiliation
{Department of Chemical Engineering, Carnegie Mellon University, 15213, USA}
\affiliation[meche]
{Department of Mechanical Engineering, Carnegie Mellon University, 15213, USA}
\alsoaffiliation[biomed]
{Department of Biomedical Engineering, Carnegie Mellon University, 15213, USA}
\alsoaffiliation[mld]
{Machine Learning Department, Carnegie Mellon University, 15213, USA}

\author{Anne Skaja Robinson}
\email{asrobins@andrew.cmu.edu}
\affiliation
{Department of Chemical Engineering, Carnegie Mellon University, 15213, USA}

\title[An \textsf{achemso} demo]
{\textbf{Characterization of  Phosphorylated Tau-Microtubule complex with Molecular Dynamics (MD) simulation}
}

\abbreviations{IR,NMR,UV}

\begin{document}

\begin{abstract}

 \noindent
Alzheimer's Disease (AD), a neurodegenerative disorder, is reported as one of the most severe health and socioeconomic problems in current public health. Tau proteins are assumed to be a crucial driving factor of AD that detach from microtubules (MT) and accumulate as neurotoxic aggregates in the brains of AD patients. Extensive experimental and computational research has observed that phosphorylation at specific tau residues enhances aggregation, but the exact mechanisms underlying this phenomenon remain unclear. In this study, we employed molecular dynamics (MD) simulations on pseudo-phosphorylated tau-MT complex (residue 199 \~{} 312), incorporating structural data from recent cryo-electron microscopy studies. Simulation results have revealed altered tau conformations after applying pseudo-phosphorylation. Additionally, root-mean-square deviation (RMSD) analyses and dimensionality reduction of dihedral angles revealed key residues responsible for these conformational shifts.

\end{abstract}

\section{Introduction}

\begin{figure}[t!]
    \centering
    \includegraphics[width=1.0\linewidth]{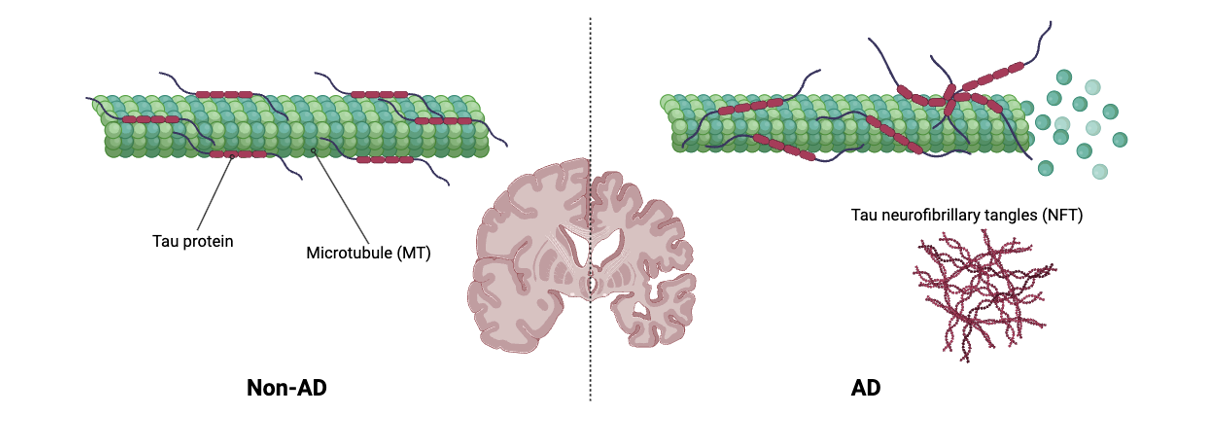}
    \caption{Comparison of the state of Tau-MT complex in AD patient brain and Non-AD. Tau proteins that stabilize the MT structures in non-AD brain become detached and form neurofibrillary tangles (NFT) in AD.}
    \label{fig:fig1}
\end{figure}

Alzheimer Disease (AD) is one of the most severe neuro degenerative disease that accompany dementia, cognitive decline and psychosis. Considering the continuous increase in human life expectancy and the number of elder population, AD has become one of the most severe health and socioeconomic problem\cite{Scheltens2021AD, Hu2014AD,Breijyeh2020adcause}. Although  the exact cause of AD still remains elusive, the deformation and aggregation of tau protein is sought as one of the causes of the AD\cite{Breijyeh2020adcause}. Tau protein stabilizes the Microtubules (MT), which are cytoskeletons that support the structure of neurons. It is unquestionable that we need tau to retain neurons in our brain\cite{AVILA2004whatisTau}. However, the problem occurs as these tau get detached from the MT and form aggregates as depicted in Figure \ref{fig:fig1}. With the absence of tau, MT would not be able to sustain its structure, unstabilizing the neuron. Additionally, the detached taus are prone to form aggregates due to their atomic force, leading to neurotoxic tau neurofibrillary tangles (NFT)\cite{Uddin2020Tauaggre}. Thus, various experimental and computational studies have been investigated to elucidate the mechanisms of the detachment of tau proteins from MT and the formulation of tau aggregates. Interestingly, among more than 50 studies comparing AD patients with controls, most of them have revealed an increase of phosphorylated tau level by approximately 300\% \cite{Hampel2010Tauphosad}. Tau protein has a total of 79 serine and threonine phosphorylation sites, and 39 of them have been verified to be related to tau aggregation\cite{Hanger2007tauphosites}. For example, an experimental study by Sengupta et al. has estimated that when T231 and S235 phosphorylation sites were blocked, each had contributed approximately 26\% and 9\% to the overall inhibition of tau binding to MT\cite{Sengupta1998262}. Likewise, numerous studies have identified which phosphorylation sites can lead to increased NFTs. However, the sophisticated mechanism of how these phosphorylations lead to the detachment of tau from MT needs further investigation. Recently, with the advent of advanced computational and simulation techniques, multiple studies have utilized these methods to discover the kinetics of phosphorylation in Tau proteins. Especially molecular dynamics (MD) simulation, a mathematical model that predicts the time-dependent behavior of a molecular system by applying various motion and energy laws to the atomic coordinance, has been widely used. Zeb et al.\cite{Zeb2019} have utilized MD simulation to investigate the interaction between Cyclin-Dependent Kinase 5 (CDK5), a protein kinase that is essential in various neuronal functions, with tau proteins leading to multiple phosphorylated sites. Ghandi et al.\cite{Gandhi2016} ran an MD simulation of 21 amino acids long phosphorylated tau to characterize the folding process. Similarly, Liu et al.\cite{Liu2020} have applied Markov State Model analysis to the MD simulation results of the R3 fragment (residue 306 \~{} 337) of tau to identify the conformation changes in phosphorylated tau.
\begin{figure}[t!]
    \centering
    \includegraphics[width=1.0\linewidth]{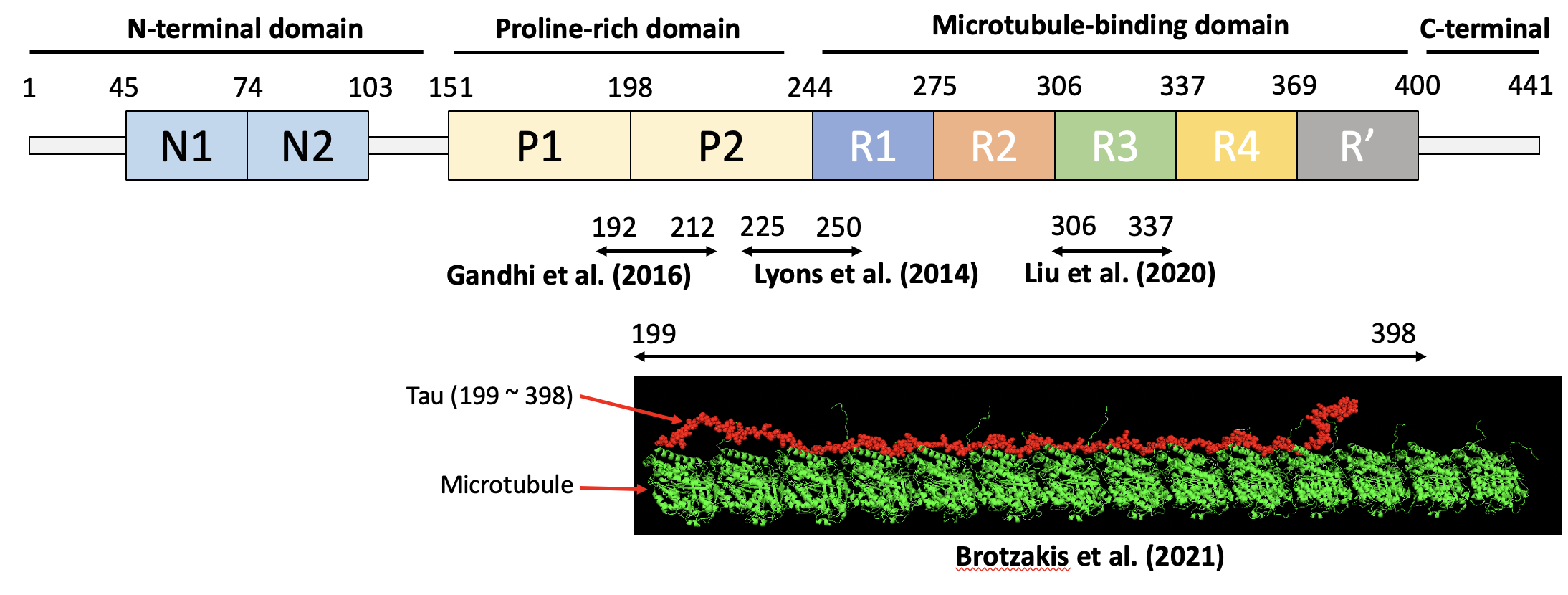}
    \caption{Regions and the corresponding residues in 2N4R tau protein. Past MD studies have mostly utilized relatively limited structural data of tau, while our study aims to characterize the entire region from P2 to R'.}
    \label{fig:fig2}
\end{figure}
However, several limitations exist in the aforementioned studies, such as using only a small portion of the Tau protein for the MD simulation. As depicted in Figure \ref{fig:fig2}, the longest isoform tau protein (2N4R) consists of several domains with a total of 441 residues. N domains refer to the N-terminal domains, and P domains are the proline-rich domains. R1 through R4 domains are the microtubule binding repeated domains. Although past works indeed suggest that the detachment of phosphorylated tau is most likely to be derived by the changes in the R1 \~{} R3 domains\cite{Zabik2017r2}; applying MD simulation to only a limited part of the tau might not provide a holistic investigation of the overall deformation of tau. Additionally, previous studies have focused exclusively on the tau protein and did not consider the interaction between the MT. To thoroughly investigate the mechanism of phosphorylated tau protein in AD, the detachment of tau from the MT must also be investigated. One of the crucial reasons for these limitations was the insufficient data of the 3D crystalized structure of complete Tau protein and MT available. However, a study in 2018 utilized cryo-electron microscopy (cryo-EM) and Rosetta remodel\cite{Baker2011rosetta} to produce the electron density map of 3D Tau-MT complex, providing a valuable resource to investigate the interactions between Tau and MT\cite{Kellogg2018cryoem}. Brotzakis et al. went further by utilizing the electron density map of Tau-MT from the former study and have constructed a structure of Tau-MT complex having P2 to R' region (residue 202 \~{} 395, PDB: 7QPC) of Tau and seven $\alpha$ - $\beta$ MT dimers\cite{Brotzakis2021emmi}. The group has produced an  MD simulation of the structure followed by the analysis of the average number of residue-MT contact and the conformational heterogeneity residues that are possible phosphorylation sites. The identified residues, such as S262 and S235 that greatly alter the stability of Tau-MT have aligned with the experimental results. In our study, going further, we aim to produce multiple MD simulations of various phosphorylated Tau-MT complexes to elucidate the detachment of Tau from the MT. Utilizing the 3D structure data from the study of Brotzakis et al., pseudo-phosphorylated Tau protein variants are created, followed by running MD simulations of each variant. pseudo-phosphorylation is a technique to mimic the effects of phosphorylation without actually adding a phosphate group by substituting the residue to negatively charged amino acids such as aspartic acid(D) and glutamic acid(E)\cite{Prokopovich2017psuedo}. Cryo-EM MetaInference (EMMI) simulation, also the method that Brotzakis et al. employed, was chosen for our study, aiming to produce phosphorylated tau trajectories in identical conditions and to compare with the WT tau's result from their study. The simulation results will be analyzed by identifying their properties and utilizing dimensionality reduction methods to elucidate the effect of phosphorylation.

\section{Methods}

\begin{figure}[t!]
    \centering
    \includegraphics[width=1.0\linewidth]{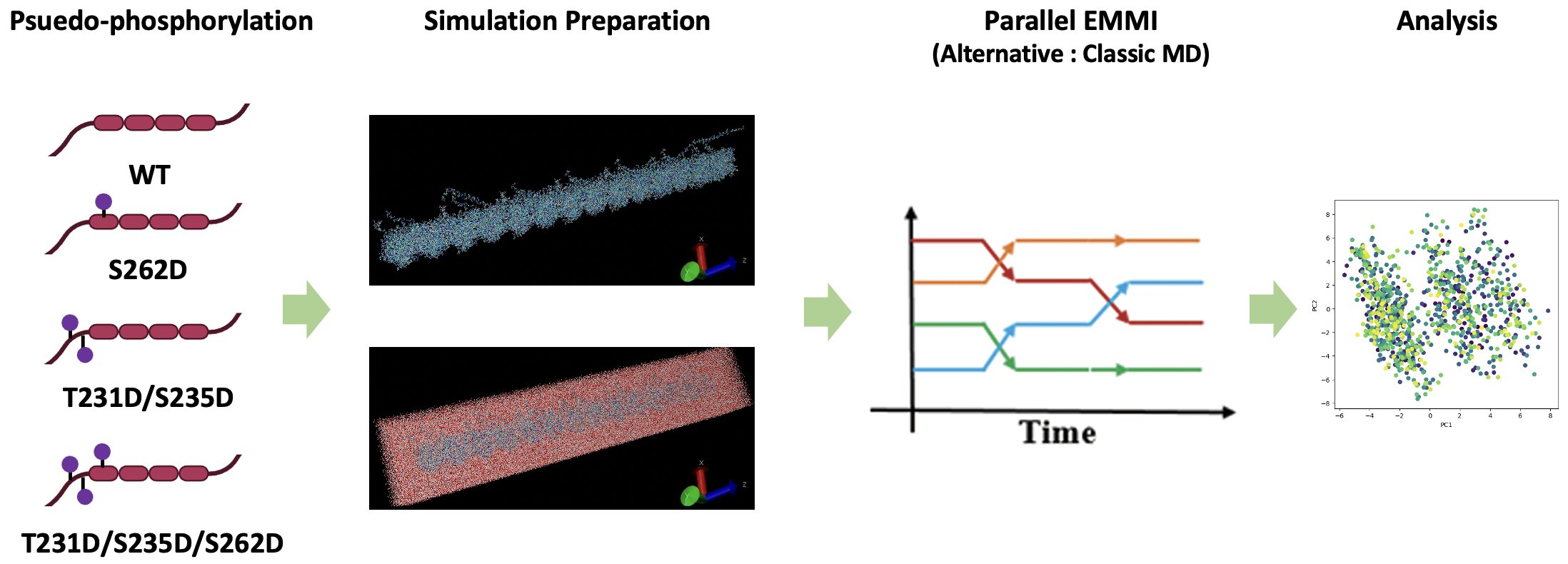}
    \caption{The overall process of the study. Three different pseudo-phosphorylated tau variants are prepared. After simulation preparation, MD simulation is applied to each variant, followed by RMSD and dimensionality reduction analysis.}
    \label{fig:fig3}
\end{figure}

\subsection{Preparing pseudo-phosphorylated tau variants}

Determining what kind of phosphorylated tau, in other words, which residue to mutate, was one of the most crucial decisions determining the overall direction of this study. We initially chose S262 and T231/S235 (mutated simultaneously) for our target. The criterion for these target sites was the location, which must be within the protein: 7QPC (residue 202 \~{} 395), and confirmed by previous phosphorylation studies that it contributes to the increment of tau aggregation. For example, Song et al. have revealed that S262D pseudo-phosphorylated tau shows significantly enhanced tau release and intracellular accumulation of tau aggregates compared to wildtype (WT)\cite{Robinson2022s262}, which aligns with multiple other studies \cite{Sengupta1998262, Brotzakis2021emmi, Hasan2021phos}. Similarly, when tau was phosphorylated in T231 and S235, its binding to MT was inhibited by 54\% and 71\% each\cite{Sengupta1998262}. Studies also mention that tau deposits isolated from AD patients often contain both phosphorylated T231 and S235\cite{Hanger2011} and are studied by pair-mutation in several works \cite{Schwalbe2015S235, Ikura1998s235}. We have utilized PyMOL software's\cite{pymol} mutagenesis function to create three variants of PDB:7QPC, resulting in S262D, T231D/S235D, and S262D/T231D/T235D tau-MT model. The last variant, S262D/T231D/S235D, was added to investigate the effect when both phosphorylation sites were employed. 

\subsection{EMMI}

Metainference\cite{Bonomi2016meta} is an advanced MD simulation method that enables the incorporation of structural data. These can be experimental data such as nuclear magnetic resonance (NMR), X-ray, fluorescence resonance energy transfer (FRET), and cryo-EM data. EMMI simulation\cite{Bonomi2018emmiorigin} is a metainference method that allows the structural protein ensemble to constantly minimize its discrepancy compared to the cryo-EM electron density map.

\begin{equation}
E_{\text{EMMI}} = E_{\text{MD}} + k_B T \sum_{r,i} \log \left[ \frac{1}{2(\overline{v}_{\text{DD},i} - \overline{v}_{\text{MD},i})}  \text{erf} \left( \frac{\overline{v}_{\text{DD},i} - \overline{v}_{\text{MD},i}}{\sqrt{2} \sigma^{\text{SEM}}_{r,i}} \right)\right]
\end{equation}

\noindent
From the above energy function, $E_{EMMI}$ refers to the total energy of the system, in this case, the protein in solvent. $E_{MD}$ is the energy from raw molecular mechanics force field energy without incorporating the experimental data. The second term refers to the metainference constraints from the cryo-EM data, represented as $i$ Gaussian mixture models (GMM). $k_B$ is the Boltzmann constant and T is the system temperature. The term inside the logarithm is the 'posterior probability.' The term posterior probability refers to the probability of observing the current state when prior information (experimental data) is acting as a constraint. Specifically, $\overline{v}_{\text{DD},i}$ is the probability for the overlap of the $i$th GMM with the entire cryo-EM GMM and $\overline{v}_{\text{MD},i}$ is the same for $i$th GMM with the GMM of current MD state while $\sigma^{\text{SEM}}_{r,i}$ representing the statistical error in calculating the ensemble average. In this way, it constantly compares the cryo-EM data with the GMM of the current state, minimizing the gap between the simulation and the experimental data. In order to deal with the heterogeneity of the system, EMMI is applied in parallel to multiple replicas (the $r$ in the second term). For this simulation, Groningen Machine for Chemical Simulations (GROMACS)\cite{Abraham2015gromacs} 2022.5 patched with PLUMED\cite{Tribello2014plumed} 2.8.3 is used in Ubuntu 22.04 Linux environment. The system hardware contains an Intel Core i7-8086K CPU and NVIDIA GeForce GTX 2080 Ti 12 core GPU with 12GB of memory.
 
\subsection{EMMI simulation}

AMBER99SB-ILDN\cite{Lindorff2010amber} force field and with TIP3P\cite{Jorgensen1983tip3p} water model is used to the simulation box of 10.5 * 12.5 * 67.0 $nm^3$. Subsequently, it was solvated with water molecules and neutralized by adding sodium and chloride ions, creating an environment of 6451 protein residues, 255884 water molecules, and 325 Ions. Energy Minimization (EM) is performed with the steepest descent minimization with a step size of 0.002 ps. The neighbor list is updated every ten steps, and long-range electrostatic interactions are calculated with the Particle Mesh Ewald (PME) method. NPT equilibrium simulation of 300 ps is performed. Leap-frog integrator is used with velocity-rescale (v-rescale) thermostat\cite{Bussi2007vrescale} and Parrinello-Rahman barostat\cite{Parrinello1981barostat}. Only bonds involving hydrogen atoms (h-bond) are constrained with LINCS\cite{hess1997lincs}. The non-bonded interactions are calculated with Verlet Scheme having a 1.0 nm cutoff for short-range interactions. PME is used as well with 0.12 nm Fourier spacing. Initial velocity is generated at temperature 310K. NVT simulation is continued from the last step of NPT simulation with an extended time of 2ns. Every condition was the same except for turning off the barostat and velocity generator. The complete parameters for the EM, NPT, and NVT steps are shown in Data and software availability. From the earlier NVT equilibrium step, six frames evenly spaced by the same number of time steps were selected. These frames served as the starting conformations for each replica simulation. For these parallel simulations, two GPU cores were allocated to each, and 100ns EMMI was carried out with the same parameters as the NVT simulation. The system configurations were recorded at intervals of 5 ps for subsequent analysis. To calculate the cryo-EM restraint, overlaps between the model and data GMMs were evaluated every two MD steps. This was done using neighbor lists with a cutoff of 0.01 and a refresh rate set to every 100 steps.

\subsection{Analysis of trajectories} 

The root mean squared deviation (RMSD) was analyzed for each simulation. Specifically, RMSD of the whole tau protein, P2/R1 region, and R2 region was investigated to observe the effect of phosphorylation in distinct regions. Next, dihedral angles of the tau protein backbone of 113 residues (residue 202 \~{} 315) were calculated for each frame, and these vectors were used as representative of the conformation of tau protein in each frame. Various dimensionality reduction methods such as Principal component analysis (PCA), t-SNE (t-Distributed Stochastic Neighbor Embedding), and UMAP (Uniform Manifold Approximation and Projection) were employed on these vectors to visualize if phosphorylation has caused conformation changes. PCA\cite{pca1,pca2} is a statistical procedure that utilizes orthogonal transformation to convert a set of observations of possibly correlated variables into a set of values of linearly uncorrelated variables called principal components. This technique is widely used in fields such as exploratory data analysis and for making predictive models. It ranks the principal components in such a way that the first few retain most of the variation present in all of the original variables. PCA simplifies the complexity of high-dimensional data while retaining trends and patterns. t-SNE\cite{tsne} is a machine learning algorithm for visualization developed by Laurens van der Maaten and Geoffrey Hinton. It is a nonlinear dimensionality reduction technique well-suited for embedding high-dimensional data for visualization in a low-dimensional space of two or three dimensions. Specifically, t-SNE models each high-dimensional object by a two- or three-dimensional point in such a way that nearby points and dissimilar objects model similar objects are modeled by distant points with high probability. UMAP\cite{umap1, umap2} is a relatively new dimensionality reduction technique that is particularly effective for visualizing clusters or groups of data points in high-dimensional data. Unlike PCA, a linear algorithm, UMAP is non-linear, making it more suitable for complex data. UMAP works by constructing a high-dimensional graph representation of the data and then optimizes a low-dimensional graph to be as structurally similar as possible. This method is often preferred for its balance between preserving of the global and local structure of data and for its computational efficiency.

\section{Result}

\subsection{EMMI of S262D variant}

\begin{figure}[t!]
    \centering
    \includegraphics[width=0.9\linewidth]{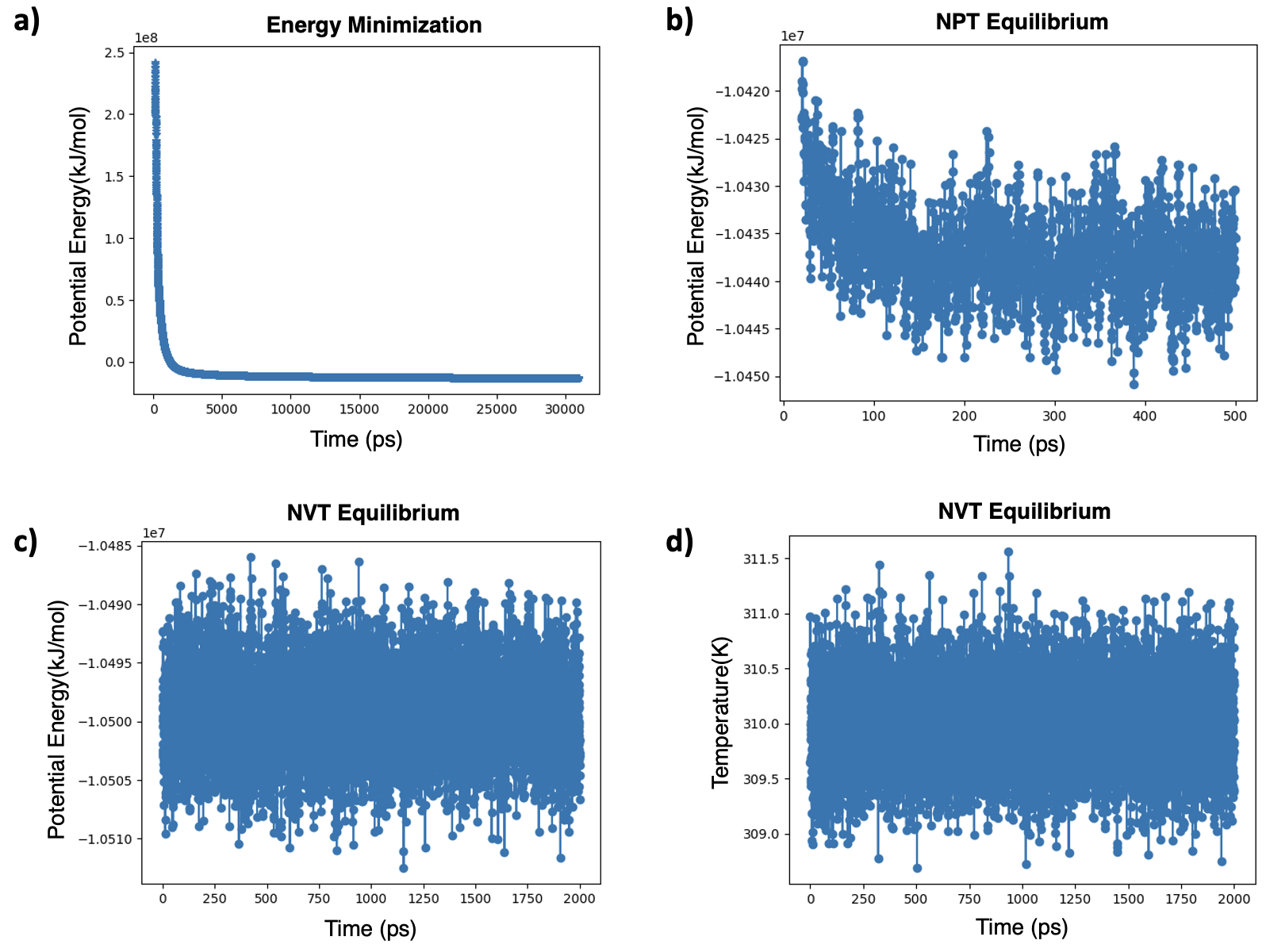}
    \caption{Results of EMMI simulation preparation on 10.5 * 12.5 * 67.0 $nm^3$ box environment. (a) Through the Energy Minimization step, the system has reached its minimum energy. (b) NPT equilibrium step and (c),(d) NVT Equilibrium step has relocated the system to a stable equilibrium state.}
    \label{fig:emmi}
\end{figure}

Initially, the solvation box size for the system was optimized. The box size should be a manageable size that the protein does not collides with the boundary but, at the same time, should avoid excessive spacing to minimize the computational costs coming from additional water molecules. Considering that Brotzakis et al. has worked with 9.7  * 11.4 * 63.6 $nm^3$box (202493 water molecules)\cite{Brotzakis2021emmi}  we started from a smaller size box of 8.0 * 13.0 * 65.0 $nm^3$(191409 water molecules). However, in this environment, the system has failed to reach equilibrium, and several residues of the $\alpha$ and $\beta$ tubulin have penetrated the boundaries of the solvation box. This behavior was still observed after increasing the box size to  9.7 * 13.0 * 65.0 $nm^3$ (236659 water molecules), even with different simulation parameter settings. Finally, the box size of 10.5 * 12.5 * 67.0 $nm^3$ (255824 water) has shown successful energy minimization and equilibrium as shown in Figure \ref{fig:emmi}. Requiring a bigger box size than the WT from the study of Brotzakis et al. might suggest that the S262D phosphorylation has destabilized the tau's attachment, leading to an increase in its deviation. After the energy minimization step, the lowest energy of the system has reached its minimum energy of -1.31e+07 kJ/mol. For the NPT equilibrium step (Figure \ref{fig:emmi} (b)), the simulation speed was 0.8ns/day, taking 14hr 59min for 0.5ns simulation. In the case of NVT equilibrium step (Figure \ref{fig:emmi}(c),(d)), the simulation speed was 0.754ns/day, taking 2d 15hr 37min for 2ns simulation. After both equilibrium processes, the system has reached its stable energy and temperature prepared for the EMMI simulation. Six conformations after every 400ns were extracted (0, 400, 800, 1200, 1600, 2000), representing the initial conformation of each replica EMMI. GMM representing the cryo-EM density map of tau-MT from previous work\cite{Brotzakis2021emmi} was utilized. Assigning 2 GPU cores to each replica, the non-bonded interactions, such as electrostatic and van der Waals interactions, are allocated to GPUs, while bonded interactions are allocated to CPUs. However, an unexpected problem was encountered. The EMMI production speed was 0.537ns/day, and it was estimated that 186.2 days are needed to produce a complete 100ns EMMI simulation. Even though we choose to use the minimum number of only two replicas, allocating 6 GPU cores to each simulation will still take 62.1 days. Several strategies could have been chosen in this situation. One was utilizing the CMU Arjuna cluster, which has 1,793 CPU and 112 GPU cores, to gain extra computing speed. Another strategy is to choose a lower cost simulation method (Classical MD simulation), downgrading the computational cost of the simulation. The former choice would bring more sophisticated results while having the risk of being unable to complete the project if additional unexpected problems arise. The latter choice can give the results much faster but has the risk of biased simulation results. Considering that the remaining time was about two to three months at the moment (Early September), we decided that doing additional experiments with the EMMI simulation has some risk of failing to complete the project in time.

\subsection{Classical MD simulation}

\begin{figure}[t!]
    \centering
    \includegraphics[width=1.0\linewidth]{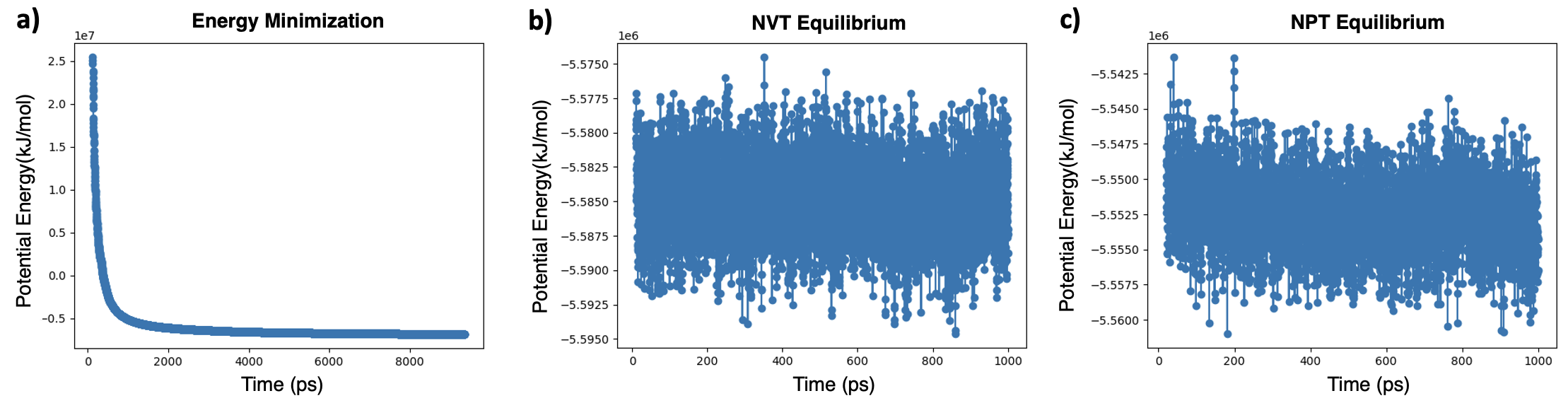}
    \caption{The Result of Classical simultion of the S262D variant. (a) Energy Minimization step, (b) NVT Equilibrium and (c) NPT equilibrium have resulted successful stabilization of the system.}
    \label{fig:normal}
\end{figure}

The tau-MT system was truncated before applying classical MD simulation to minimize unnecessary computational costs. The R3 to R' region of the tau has been removed initially. The rationale behind this decision was from previous studies on tau phosphorylation\cite{Zabik2017r2, man2023r2} reporting that the conformation changes due to tau phosphorylation are mainly found in the remaining R1 and R2 regions. Accordingly, three $\alpha$ and four $\beta$ dimers are truncated as well leaving a system of four $\alpha$, three $\beta$ dimers, and 113 tau protein residues of P2 to R2 region (residue 199 \~{} 311) . As a result, the system has decreased to a box size of 9.7 * 13.0 * 65.0 $nm^3$ (236659 water molecules). The same parameters were used in energy minimization, NVT equilibrium, and NPT equilibrium steps as in the EMMI simulation. For the S262D variant, after the energy minimization step, the truncated system has reached its minimum of -6.83e+06 kJ/mol after 9382 steps (Figure \ref{fig:normal}(a)). The NVT equilibrium (Figure \ref{fig:normal}(b)) took 1hr 35 mins at the rate of 15.037 ns/day while the NPT equilibrium (Figure \ref{fig:normal}(c)) took 1hr 36min at the rate 14.880 nd/day. The T231D/S235D and T231D/S235D/S262D were all conducted with the same parameters and procedure, resulting in similar outcomes. After such procedures, the main classical MD simulation speed was increased to 15.902 ns/day, completing a 100ns simulation in 6 days, 6 hours, and 55 mins. The time step was 0.002 ps, requiring 50 million steps for a 100ns trajectory. Conformation coordinates were saved after 50,000 steps, obtaining 1000 frame trajectories. Thus, in this way, we have acquired four 100ns MD simulation trajectories of WT, S262D, T231D/S235D, and S262D/T231D/S235D.

\subsection{RMSD (P2,R1,R2)}

\begin{figure}[t!]
    \centering
    \includegraphics[width=1.0\linewidth]{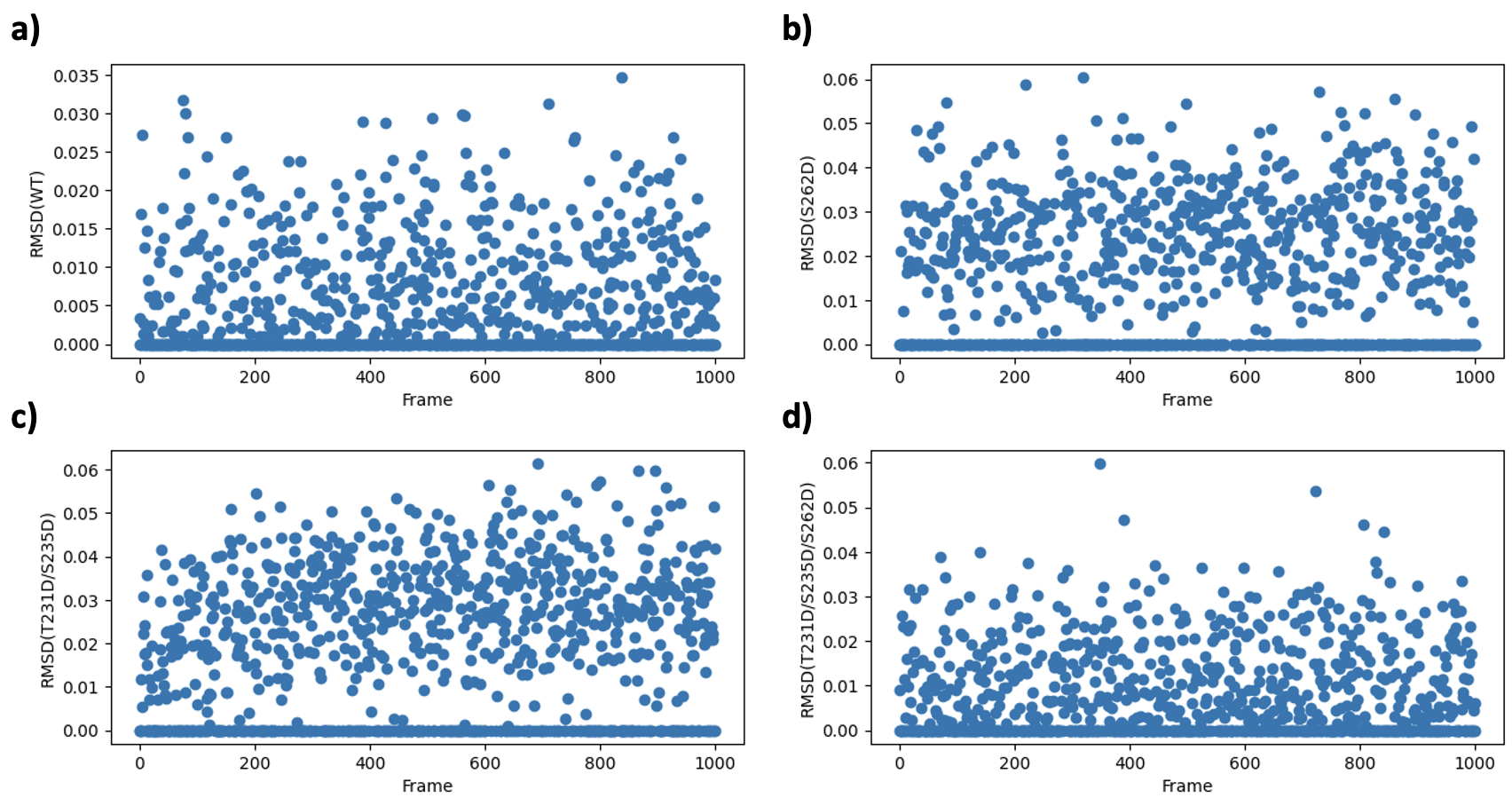}
    \caption{RMSD of tau protein from P2 to R2 regions (residue 199 \~{} 311) of (a) WT, (b) S262D variant, (c) T231D/S235D variant, and (d) T2312D/S235D/S262D variant. All RMSD are in unit of nm. }
    \label{fig:rmsd_all}
\end{figure}

\begin{table}[t!]
\begin{tabular}{|l|c|c|c|}
\hline
                               & Max    & Average & \multicolumn{1}{l|}{Standard Dev} \\ \hline
    \textbf{WT}                & 0.0347 & 0.0049  & 0.0070                            \\
    \textbf{S262D}             & 0.0605 & 0.0152  & 0.0151                            \\
    \textbf{T231D/S235D}       & 0.0614 & 0.0184  & 0.0167                            \\
    \textbf{T231D/S235D/S262D} & 0.0599 & 0.0074  & 0.0100                            \\ \hline
    \end{tabular}
    \caption{\label{tab:acc1}Maximum, average and strandard deviation values for RMSD of tau proteins from P2 to R2 regions throughout the trajectory (1000 frames). All values are in unit of nm.}
\end{table}

Figure \ref{fig:rmsd_all} shows the RMSD from the entire P2 to R2 regions (residue 199 \~{} 311) for WT and each variant, while Table \ref{tab:acc1} displays the maximum, average and standard deviation of the RMSD of each variant. From the result, we can observe that the S262D and T231D/S235D variants show an increase in the average and standard deviation after applying pseudo-phosphorylation to these positions. This result might not necessarily indicate that the pseudo-phosphorylation has led to the detachment of tau from the MT but can indicate meaningful conformation changes and increased flexibility or movement within certain parts of the protein. Interestingly, applying both T231D/S235D and S262D showed less deviation from the WT than each individual mutation, which was unexpected. Thus, we have analyzed the RMSD of separate P2/R1 and R2 regions for further investigation.

\subsection{Seperate RMSD (P2/R1 and R2)}

\begin{figure}[t!]
    \centering
    \includegraphics[width=1.0\linewidth]{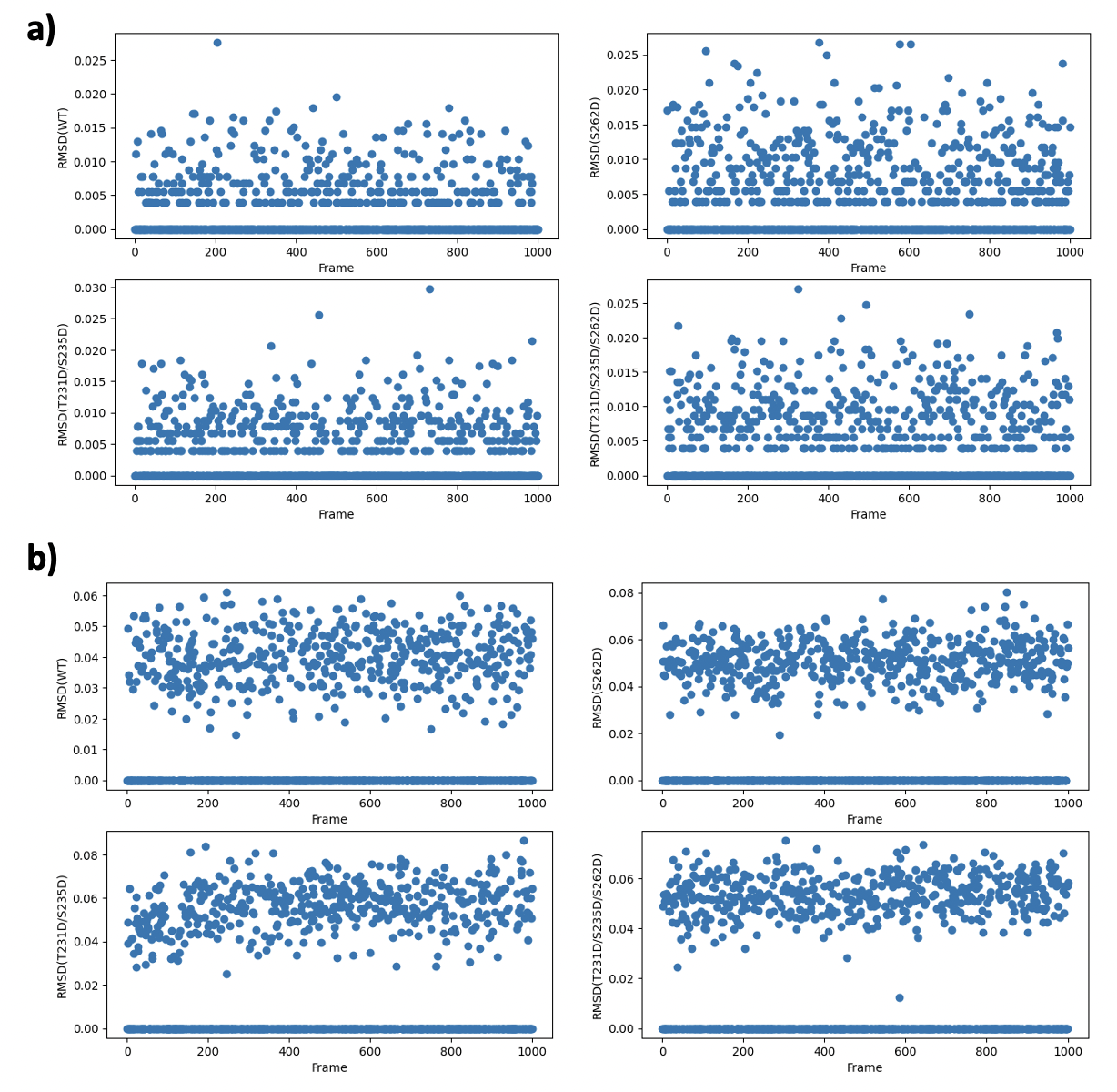}
    \caption{RMSD of tau protein  (a) P2/R1 (residue 199 \~{} 275) and (b) R2 region (residue 276 \~{} 311). All RMSD are in unit of nm.}
    \label{fig:rmsd_p2r1_r2}
\end{figure}

\begin{table}[]
\begin{tabular}{|l|ccc|ccc|}
\hline
                                    & \multicolumn{3}{c|}{\textbf{P2 + R1}}                                 & \multicolumn{3}{c|}{\textbf{R2}}                                    \\ \cline{2-7} 
    \multicolumn{1}{|c|}{\textbf{}} & \multicolumn{1}{c|}{Max}    & \multicolumn{1}{c|}{Average} & Std    & \multicolumn{1}{c|}{Max}    & \multicolumn{1}{c|}{Average} & Std    \\ \hline
    \textbf{WT}                     & \multicolumn{1}{c|}{0.0276} & \multicolumn{1}{c|}{0.0024}  & 0.0042 & \multicolumn{1}{c|}{0.0611} & \multicolumn{1}{c|}{0.0190}  & 0.0210 \\
    \textbf{S262D}                  & \multicolumn{1}{c|}{0.0268} & \multicolumn{1}{c|}{0.0043}  & 0.0058 & \multicolumn{1}{c|}{0.0802} & \multicolumn{1}{c|}{0.0263}  & 0.0262 \\
    \textbf{T231D/S235D}            & \multicolumn{1}{c|}{0.0297} & \multicolumn{1}{c|}{0.0033}  & 0.0047 & \multicolumn{1}{c|}{0.0866} & \multicolumn{1}{c|}{0.0262}  & 0.0289 \\
    \textbf{T231D/S235D/S262D}      & \multicolumn{1}{c|}{0.0271} & \multicolumn{1}{c|}{0.0041}  & 0.0055 & \multicolumn{1}{c|}{0.0753} & \multicolumn{1}{c|}{0.0258}  & 0.0273 \\ \hline
    \end{tabular}
    \caption{\label{tab:acc2}Maximum, average and strandard deviation values for RMSD of tau proteins from P2/R1 and R2 region throughout the trajectory (1000 frames). All values are in unit of nm. }
\end{table}

Figure \ref{fig:rmsd_p2r1_r2} depicts the RMSD of P2/R1 (residue 199 \~{} 275) and R2 (residue 276 \~{} 311). Table \ref{tab:acc2} shows each region's maximum, average, and standard deviation of the RMSD. We can see from the result that the pseudo-phosphorylation did not significantly affect the RMSD of the P2/R1 region, as the properties of the mutated tau's RMSD do not have a noticeable deviation compared to the wild type. However, the RMSD in the R2 region was different, showing a remarkable increase in maximum, average, and standard deviation. Even T231D/S235D/S262D, which have shown less deviation than S262D and T231D/S235D mutation in the whole region (P2 \~{} R2), showed a noticeable increased deviation in the R2 region. It is intriguing that although the pseudo-phosphorylation is made in the P2 (T231D/S235D) and R1 (S262D) regions, both have affected the R2 region more than the mutation location. The R2 region has been reported multiple times as the core region of tau aggregation in AD. A study in 2014 utilized Single molecular Forster resonance energy transfer (smFRET) to WT tau and tubulin heterodimers and revealed that the R2 region was identified as a key sequence in tau aggregation due to their propensity for $\beta$ sheet formation\cite{Elbaum-Garfinkle2014r21}. Also, part of the R2 region (residue 273 \~{} 284) was analyzed by MD simulation in 2015 and reported that it was highly unstructured with interconverting conformers\cite{Ganguly2015r22}. To further investigate the observed deviation and R2 region, we applied various dimensionality reduction methods to each phosphorylated system.

\subsection{PCA}

\begin{figure}[t!]
    \centering
    \includegraphics[width=0.8\linewidth]{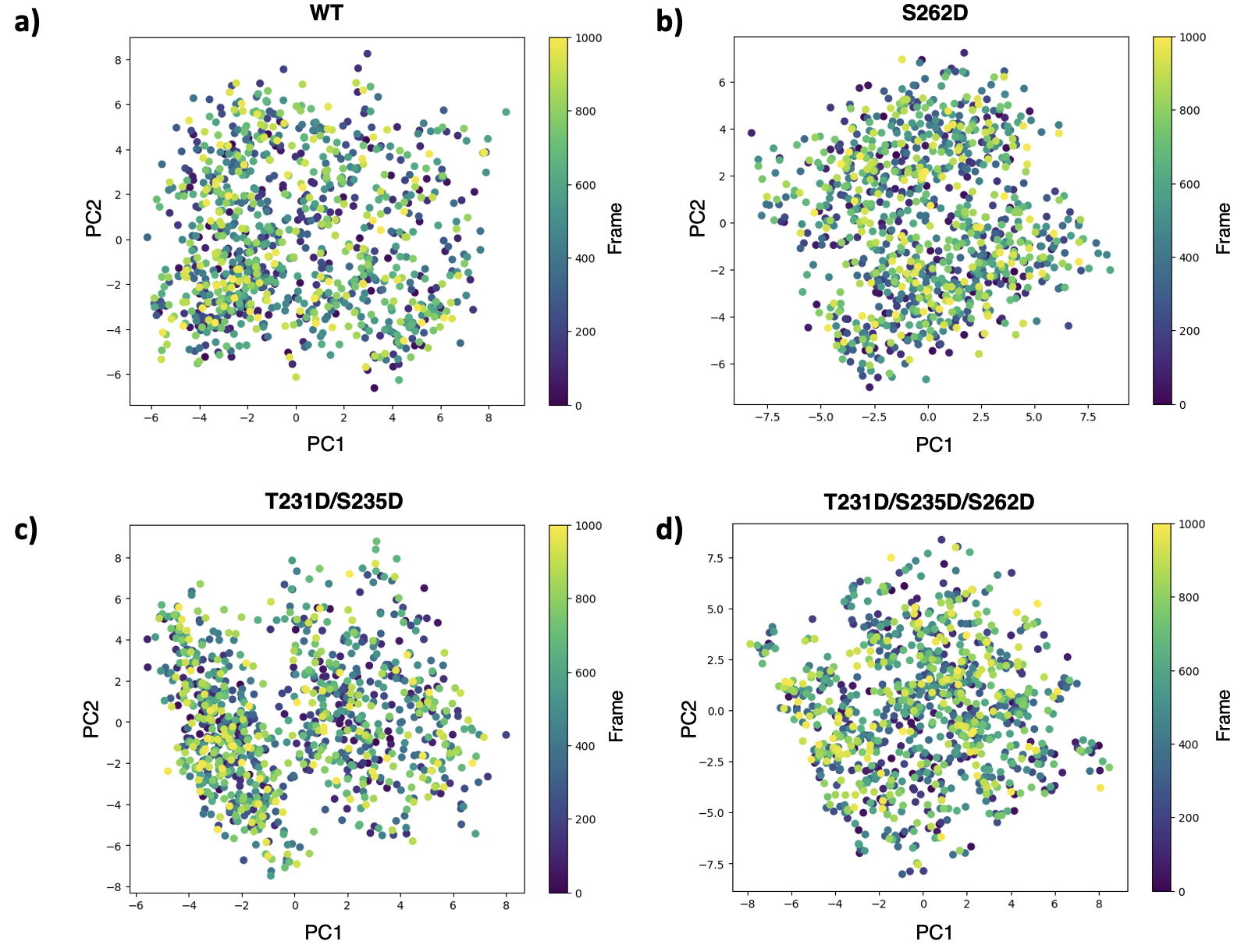}
    \caption{PCA result of (a) WT, (b) S262D, (c) T231D/S235D and (d) T231D/S235D/S262D. Dihedral angles of tau backbone atoms are extracted from each frame and converted into 2 dimension. }
    \label{fig:pca_all}
\end{figure}

Figure \ref{fig:pca_all} shows the PCA result of each variant. The $\phi$ and $\psi$ dihedral angles of each tau residue are calculated, resulting in a total length of 226 (113 * 2) vectors, representing the tau conformation in each frame. These vectors are extracted from each frame, and the trajectory are represented as a 226 * 1000 (number of frames) matrix. This matrix was reduced to 2*1000 in size through PCA and visualized as above. Noticeably, we can observe that the frames are clustered into upper and lower regions in the S262D variant (Figure \ref{fig:pca_all}(b))whereas it is clustered into right and left in T231D/S235D variant (Figure \ref{fig:pca_all}(c)). Compared with the PCA of WT (Figure \ref{fig:pca_all}(a)), which is absent of any bias, we can conclude that adding pseudo-phosphorylation has resulted in the advent of a different conformation of tau. Unexpectedly, the triple mutation of T231D/S235D/S262D showed no clustering which will be discussed in following analyses.

\subsection{Intergration of RMSD to dimensionality reduction methods}

\begin{figure}[t!]
    \centering
    \includegraphics[width=0.8\linewidth]{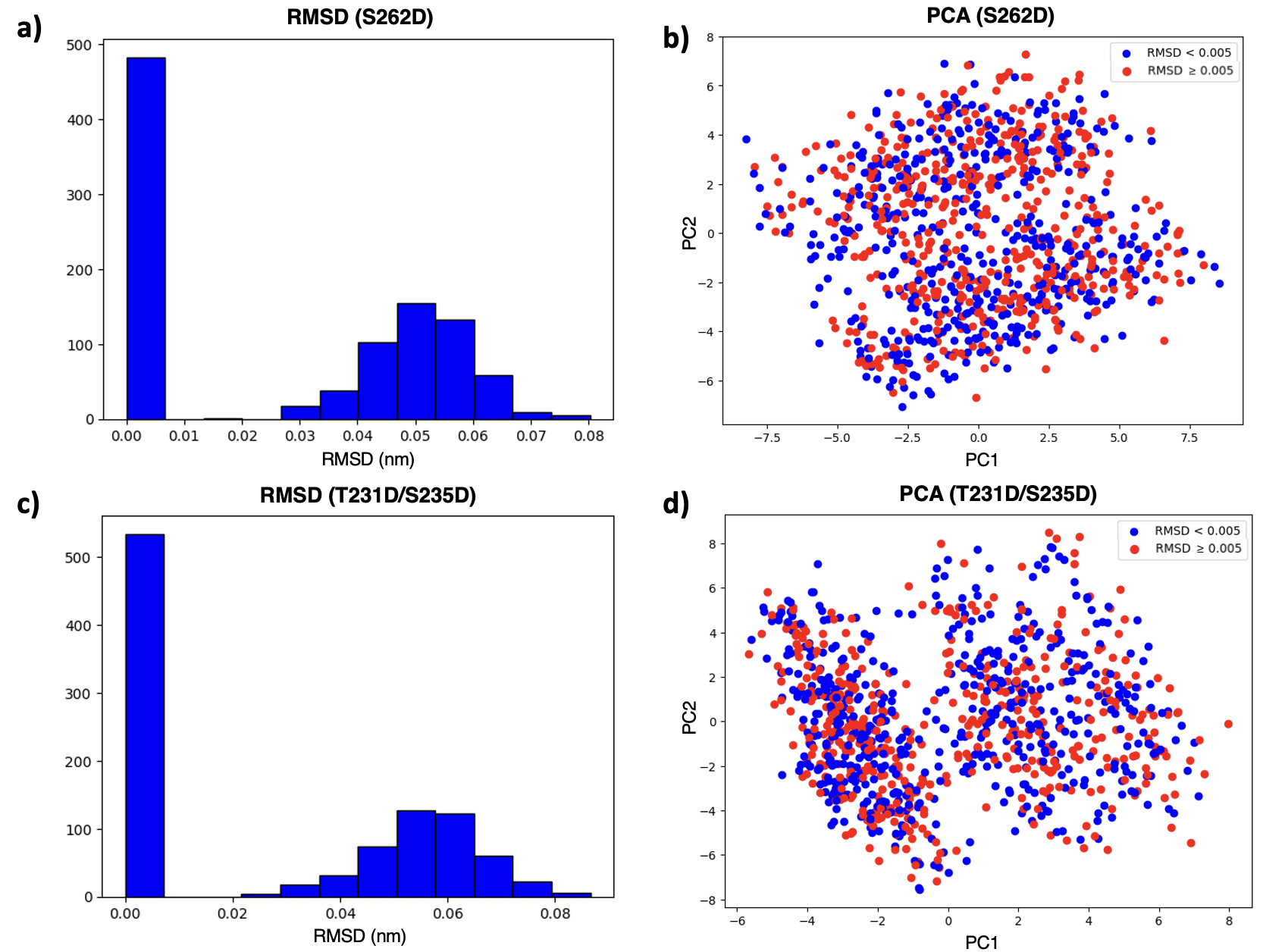}
    \caption{Integration of RMSD to PCA results. Histogram (a) shows the R2 region RMSD values distribution of the S262D variant trajectory. The PCA of S262D is labeled by R2 region RMSD value, marking the frames with lower RMSD than 0.005 as blue and higher than 0.005 as red. (c) and (d) shows the corresponding histogram and labeled PCA for T231/S235D.}
    \label{fig:hist+pca}
\end{figure}

\begin{figure}[t!]
    \centering
    \includegraphics[width=0.8\linewidth]{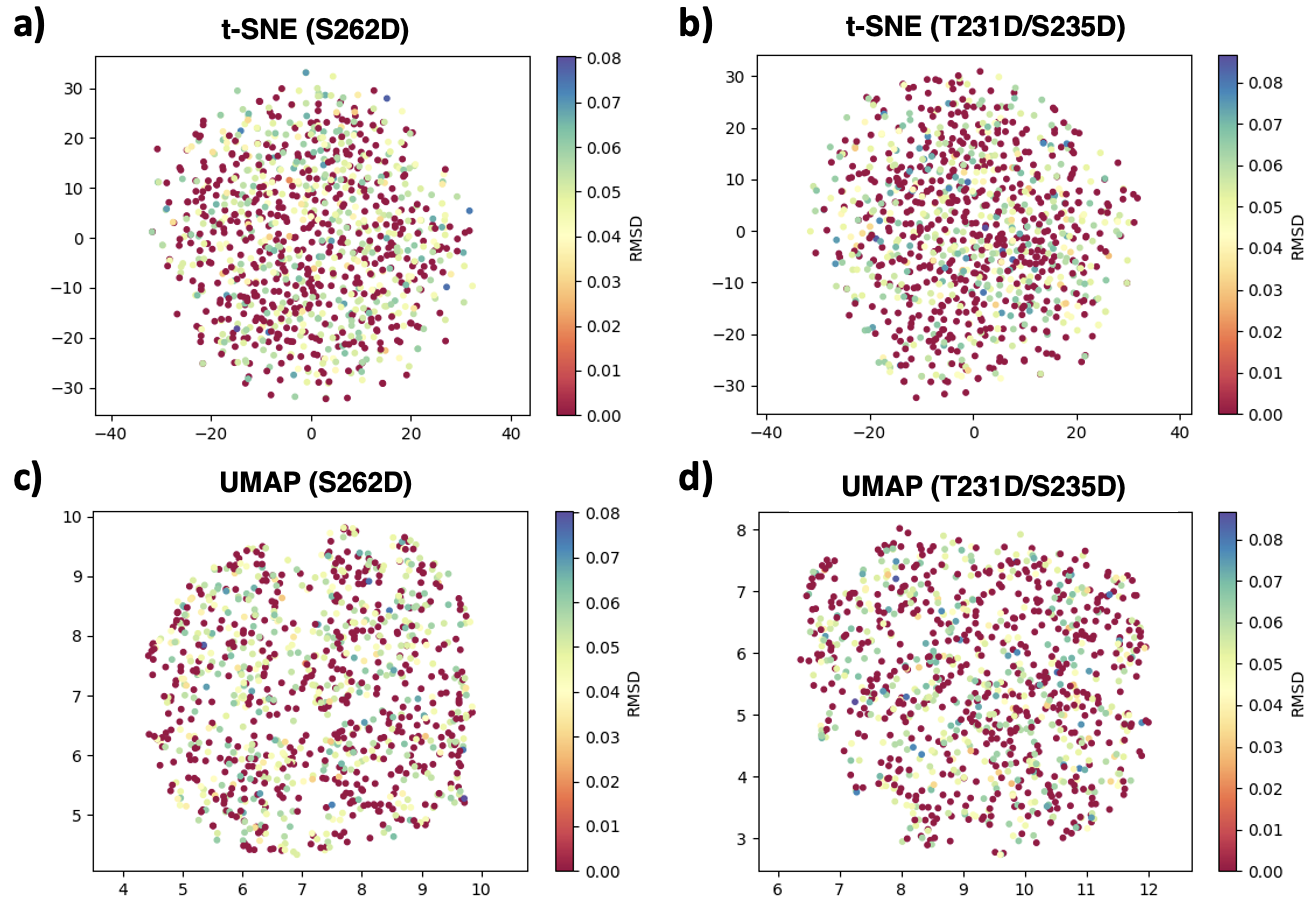}
    \caption{The t-SNE and UMAP result of S262D and T231D/S235D variant. (a) is the t-SNE plot of S262D variant labeled by the RMSD values of the R2 region and (b) is the same for T231D/S235D variant. (c) and (d) are the corresponding UMAP result labeled by the R2 region RMSD}
    \label{fig:tsne_umap}
\end{figure}

For additional analysis, we have integrated RMSD with various dimensionality reduction methods. First, we wanted to investigate if the clusters in previous PCA results have correlations with the RMSD values of the R2 region. Since the clustering in PCA and increased deviation of the R2 region RMSD result from the introduction of pseudo-phosphorylation, we expected them to be related. After visualizing the distribution of R2 region RMSD values (Figure \ref{fig:hist+pca} (a), (c)), we have observed that approximately half (400 \~{} 500) of the frames have RMSD values less than 0.005 nm and the other half is distributed in a Gaussian curve between 0.02 nm and 0.08 nm. Thus, we assumed there is a chance that these two groups correspond to the two clustered groups in PCA. However, as observed in Figure \ref{fig:hist+pca} (b) and (d), both groups of frames that have the RMSD larger than 0.005 are evenly distributed in both PCA clusters, depicting that R2 region RMSD values do not represent the clusters in PCA. Next, we assumed the PCA method was not sophisticated enough to retain the necessary information in the dihedral vectors. Thus, we have repeated the above analysis utilizing t-SNE and UMAP. t-SNE and UMAP models are capable of understanding the non-linearity of the data, while PCA can only interpret the data in linear relationships. Also, they are more effective in preserving the local structure of the data than PCA\cite{njue2020dimensionality}. Dihedral angle vectors of 226 lengths representing each frame are also used for these methods and reduced to 2D vectors. Additionally, the frames are colored according to their R2 region RMSD values to see the correlation, and the results are shown in Figure \ref{fig:tsne_umap}. Opposing our hypothesis, advanced dimensionality reduction methods also suggest no correlation between the R2 region RMSD and the tau conformations.

\subsection{PCA of R2 region}

\begin{figure}[t!]
    \centering
    \includegraphics[width=0.8\linewidth]{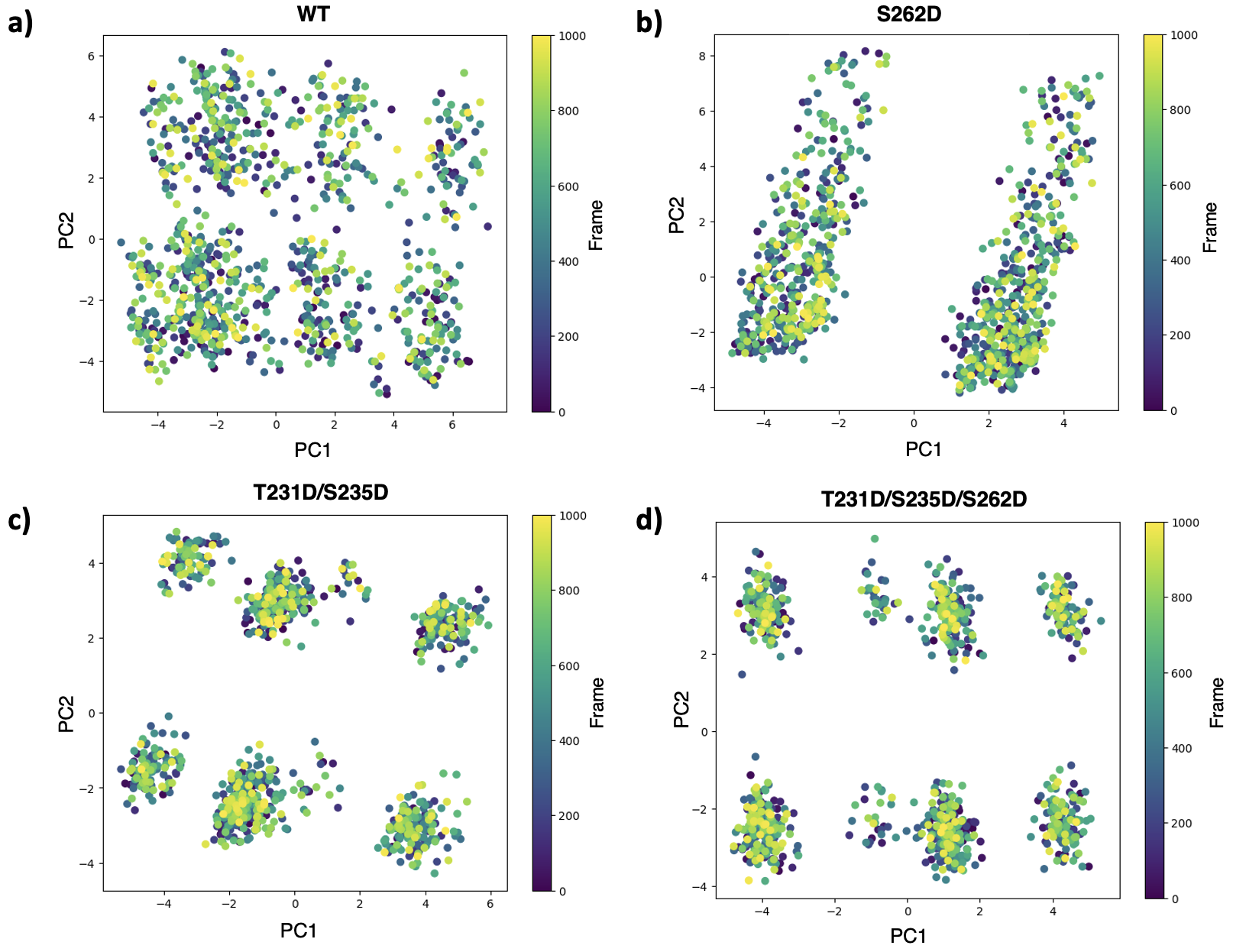}
    \caption{PCA result of (a) WT, (b) S262D, (c) T231D/S235D and (d) T231D/S235D/S262D. Dihedral angles of R2 region are extracted from each frame and converted into 2 dimension. }
    \label{fig:pca_r2}
\end{figure}

\begin{figure}[t!]
    \centering
    \includegraphics[width=1.0\linewidth]{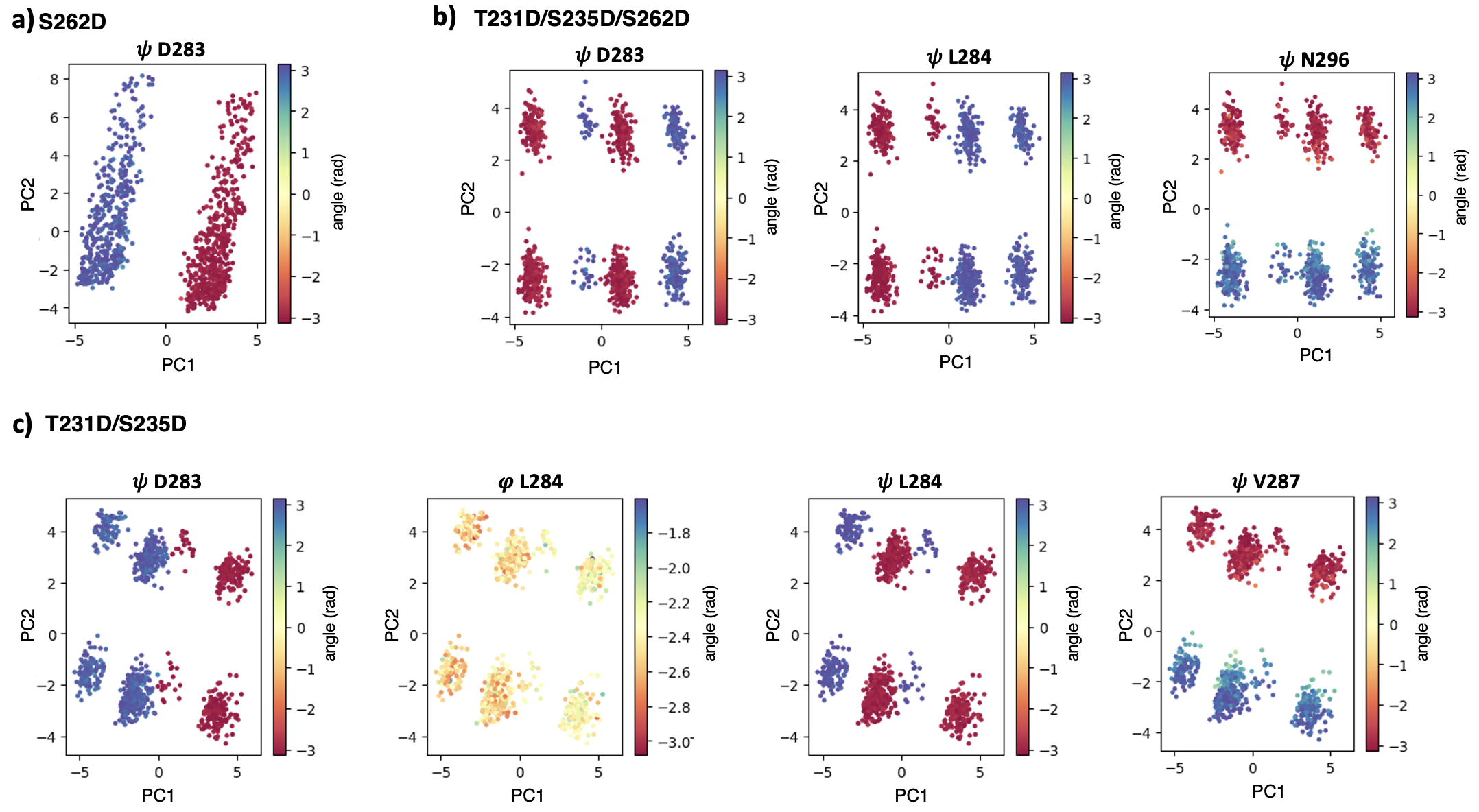}
    \caption{Angle analysis of PCA results of R2 region angles. PCA plot of (a) S262D, (b) T231D/S235D/S262D, and (c) T231D/S235D are labeled with each dihedral angles. Only angles that were able to distinguish the clusters are presented.}
    \label{fig:pca_r2_com}
\end{figure}

From previous RMSD analysis, the R2 region was dictating the overall changes in the conformation of the system. Thus, we hypothesized that removing the information on the P2/R1 region and only focusing on the R2 region might give a more robust result. Thus, only the dihedral angle of the R2 region (residue 276 \~{} 309), a total of 68 angles, are extracted from each frame and analyzed through PCA in the same method as the previous PCA. \ref{fig:pca_r2} shows the PCA result of R2 region dihedral angles. We can observe that this result shows more robust clustering than the analysis with the P2 \~{} R2 region tau. Interestingly, the T231/S235D/S262D variant, which did not show any clustering in the previous PCA analysis, is now divided into 8 clusters (Figure \ref{fig:pca_r2}). Next, to elucidate which of the 68 angles contribute to PCA clustering, we have labeled the PCA results with each angle value exclusively. While most of the angles showed uniform distribution throughout the clusters, several angles separated the PCA clusters (Figure \ref{fig:pca_r2_com}). $\psi_{D283}$ was found to be contributing to the formation of different conformation in all variants. For both the T231/S235 and T231/S235/S262D variants, $\psi_{L284}$ additionally dictated the difference of conformations. We assume that deviation of $\psi_{L284}$ might be the effect of T231/S235D double mutation. Interestingly, the $\phi_{L284}$ and $\psi_{V287}$ was observed as key factors in T231/S235D variant but was not in the T231D/S235D/S262D. No literature has yet reported the correlation between these residues and phosphorylation in tau proteins. However, we believe these results can shed light on the effect of phosphorylation on the tau-MT complex in AD and other tauopathies.

\section{Discussion}

So far, we have applied MD simulation to tau-MT complex (PDB: 7PQC) variants to investigate the effect of phosphorylation. Three kinds of pseudo-phosphorylation, S262D, T231D/S235D, and T231D/S235D/S262D are selected which are frequently reported as significant phosphorylation residues from previous literatures\cite{Robinson2022s262,Sengupta1998262, Brotzakis2021emmi, Hasan2021phos, Schwalbe2015S235, Ikura1998s235}. The variants ran for 100ns each in a protein in water environment with classical MD simulation. By analyzing the RMSD of the trajectories, it has been observed that the phosphorylated tau proteins show an increase in the deviation of RMSD compared to WT(Figure \ref{fig:rmsd_all}). By separating the region into P2/R1 and R2, we have observed that these deviations are originating from the changes in the R2 region. Although observing deviation in the R2 region after applying pseudo-phosphorylation in the P2 and R1 region has not been reported in previous MD simulation studies since the majority of them have targeted a single domain (single R1, single R2, etc.), we believe our result has meaning in connecting the significant phosphorylation sites (S262D/T231D/S235D) with the MT binding R2 region which multiple literatures have identified as one of the most crucial domains affecting tau aggregation\cite{Hasan2021phos, man2023r2, Man2023r2.2}. Furthermore, we extracted the tau protein's dihedral angles representing each frame's protein conformation and visualized this through various dimensionality reduction methods: PCA, t-SNE, and UMAP. Although t-SNE and UMAP did not detect distinct conformation groups within the trajectory, PCA results have shown that several conformation groups exist in the trajectories of pseudo-phosphorylated variants (Figure \ref{fig:pca_all}), which are more robustly observed in the exclusive PCA analysis on R2 region dihedral angles (Figure \ref{fig:pca_r2}). Through component analysis of the PCA of the R2 region (Figure \ref{fig:pca_r2_com}), we have identified several dihedral angles such as $\psi_{D283}$ and $\psi_{L284}$, that are dictating the distinct conformations. From the fact that these deviations in angles are not observed in non-phosphorylated WT, we concluded that the pseudo-phosphorylation mutations certainly derive these and can elucidate the mechanism of phosphorylation in tau protein, leading to aggregation of tau in AD. However, this study also has several limitations, considering its robustness. Unlike other advanced MD simulations, the classical MD simulation method is prone to give biased results by trapping the energy to a local minimum\cite{Bonomi2016metainferece2}. Thus, the study needs to be confirmed by running duplicates or with advanced MD simulations. Future research would run the full PDB:7PQC with EMMI simulation with additional significant phosphorylation cites such as S202 and T205\cite{Rankin2005202205}. Also, experimental confirmation of current results, such as site-directed mutagenesis in tau proteins, is needed.

\section{Conclusion}

This study has utilized classical molecular dynamics (MD) simulations of pseudo-phosphorylated tau-MT variants to explore the effects of phosphorylation. By analyzing RMSD trajectories and backbone dihedral angles through various dimensionality reduction methods, we observed significant deviations in the R2 region of phosphorylated tau proteins compared to the wild type. We identified several residues that derive these conformation changes. These findings suggest a direct impact of pseudo-phosphorylation on tau protein conformation shifts, potentially elucidating the effect of phosphorylation on tau aggregation in AD. However, the study acknowledges limitations due to the potential biases of classical MD simulation and the necessity for further validation through duplicate simulations, advanced MD methods, and experimental confirmation. Future research should expand the scope of the simulation and include additional phosphorylation sites for a more comprehensive understanding.

\section{Data and software availability}
The necessary information containing the codes and data for downstream tasks used in this study are available here: 
\href{https://github.com/Andrewkimmm/Tau-MT}{https://github.com/Andrewkimmm/Tau-MT}

\begin{acknowledgement}

This work is supported by the Carnegie Mellon University Chemical Engineering Department. 

\end{acknowledgement}

\bibliography{reference}

\end{document}